\documentclass[11pt]{article}

\usepackage[margin=1in]{geometry}
\usepackage{float}
\usepackage{adjustbox}
\usepackage{booktabs}
\usepackage{xcolor}
\usepackage{amsmath,bm}
\usepackage{amssymb}
\usepackage{caption}
\usepackage{amsthm}
\usepackage{subcaption}
\usepackage{mathtools}
\usepackage{comment}
\usepackage{soul}
\usepackage{enumitem}
\usepackage{multirow}
\usepackage{hyperref}
\usepackage{authblk}

\theoremstyle{definition}

\usepackage[
    backend=biber,
    style=apa,
    natbib=true,
    uniquename=false,
    url=false,
    isbn=false,
    doi=true,
    eprint=false,
]{biblatex}
\title{Forecasting of volatility and risk premia in electricity markets}
\author[1,2,3]{Thomas K. Kloster}
\author[2]{Fred Espen Benth}

\affil[1]{Department of Economics and Business Economics, Aarhus University}
\affil[2]{Department of Data Science and Analytics, BI Norwegian Business School}
\affil[3]{CoRE, Center for Research in Energy: Economics and Markets}

\begin{document}
\maketitle

\begin{abstract}
    We study forecasting of the realized covariation in electricity markets. The realized covariation in this context is a matrix-valued representation of the latent infinite-dimensional covariance operator and a parsimonious matrix-HAR type model is constructed to facilitate estimation. We test the model on one-week ahead forecasts of the weekly realized covariation and find that the inclusion of longer time horizons and renewable generation information adds important predictive power. We also investigate the prediction of risk premia in electricity forward markets and find that our variance forecasts provide substantially improved forecasts of spread risk premia compared to standard methods relying on backward looking volatility.
\end{abstract}

\section{Introduction}
The present paper is concerned with forecasting of the realized covariation (RCV) of electricity prices in European generation zones. Volatility forecasting has a long history in financial econometrics \citep[see, e.g., ][for overviews of the literature]{HansenLunde2005,ChristensenSiggaardVeliyev2023}, but since electricity markets function very differently from traditional equity and commodity markets, the literature on volatility estimation in these markets is still very sparse. 

A fruitful approach to volatility estimation in electricity markets has recently been proposed in \textcite{KlosterBenth2026}, which is inspired by the abstract modelling framework of \textcite{Kloster2026}. The present paper builds upon these results and the main contribution is twofold: First, we show that the realized covariation as introduced in \textcite{KlosterBenth2026} is predictable over short horizons via simple matrix-HAR inspired models. Second, we show that the forecasted realized covariation improves standard forecasts of forward risk premia. This shows that, while the realized covariation arises from somewhat abstract considerations, it carries highly useful information for market participants and analysts.

The crucial observation of \textcite{KlosterBenth2026} is that the ``price of electricity'' at a given point in time really represents the average price of electricity \emph{over a delivery period}. In this view, the true price of electricity at each instant in time is a latent object and the prices that one observe in the market are averages/integrals of the latent price. In practice, the quoted prices of electricity (spot prices) are determined each day via an auction, and we denote by $P_t^{(i)}$ the price of consuming 1MWh of electricity at any point during delivery period $i$ on day $t$. Supposing that the day is partitioned into $d$ delivery periods, the prices $P_t^{(1)},\ldots ,P_t^{(d)}$ are determined simultaneously on day $t-1$, and we thus observe spot prices as a vector-valued time series $P_t=(P_t^{(1)},\ldots,P_t^{(d)})^\top\in\mathbb{R}^d$. Since electricity can be consumed continuously over the course of the day, the latent ``true'' price process is therefore inherently infinite dimensional (think of it as a curve). Indeed, if $X_t(h)$ denotes the price of consuming 1MWh of electricity during the instant point in time $h$ on day $t$, then the observed price $P_t^{(i)}$ corresponds to 
\begin{equation}\label{eq:price_local_average}
P_t^{(i)}=\frac{1}{\lvert H_i\rvert}\int_{H_i}X_t(h)dh,
\end{equation}
where $H_i$ represents delivery period $i$ with length $\lvert H_i\rvert$. Specifying the latent price $X_t(h)$ as an evolution equation in an infinite dimensional Hilbert space (a function-valued stochastic differential equation), one can define a realized variance estimator on the observation space, which is in complete analogy to the finite dimensional setting. This corresponds to the latent price series $t\mapsto X_t(\cdot)$ being a function-valued time series, and the observed finite dimensional time series is then a particular projection of this function-valued object. 

Since the observation frequency of electricity prices is inherently low frequency (daily), the realized covariation induces two main sources of bias: one from the drift and one from mean-reversion/propagation effects. It is shown in \textcite{KlosterBenth2026} that deterministic variation from mean-reversion easily accounts for more than $40\%$ of the total variation in prices and hence a feasible plugin estimator to account for this is developed. A more subtle effect owing uniquely to the infinite dimensionality is that, even when accounting for deterministic drift and propagation effects, the RCV only identifies a certain smoothed version of the integrated variance operator, where the smoothing is directly tied to the mean-reversion mechanism of the price evolution. This means that we cannot generally identify the usual notion of integrated variance, but after proper de-biasing we still identify an effective variance of the price innovations. In fact, even in the infill limit, it is not generally possible to identify the pure integrated variance operator, as shown in \textcite{BenthSchroersVeraart2022,BenthSchroersVeraart2024}. This does not consitute a problem since the effective variance -- and by proxy the RCV -- enters into many relevant computations, such as hedge ratios and confidence bands for price forecasts in the infinite dimensional setting. It is therefore desirable to have reliable forecasts of the RCV, just as it is desirable to have reliable forecasts for the realized variance of asset prices. 

As the RCV is inherently a noisy proxy of the true integrated covariance operator, we introduce a class of models inspired by the matrix heterogeneous autoregressive (HAR) model of \textcite{MatrixHAR}. In particular, we adopt the now classical view of \textcite{Corsi2009}, where the primitive forecasting target is the estimated RCV itself and not the latent infinite dimensional covariance operator. In analogy to the classical HAR framework, we incorporate four autoregressive components, corresponding to daily, weekly, monthly, and quarterly time horizons. The quarterly component is typically not included in HAR type models for stock price volatility, but serves to better incorporate long range dependencies and seasonal effects. We show that the quarterly component improves forecasts compared to a model with only daily, weekly, and monthly horizons, and we run an extensive horserace of model extensions based on their forecasting performance in the German generation zone. Based on a model confidence set approach, we determine that the best performing models generally incorporate information on renewable production, current price level, and time-varying spillover effects in the cross sectional variance. 

We conduct several robustness checks that are very relevant for empirical applications. First, we consider how the forecasts match the realized variances of various zero-sum portfolios, which is relevant for cross sectional risk management via, e.g., spread trading or static hedging. We find that the best performing forecasting models are all well-calibrated to the cross sectional dependence structure, but that the models tend to over-predict the realized volatility. This over-prediction is an artifact of the heavily right-skewed conditional variance and represents a common structural issue in volatility forecasting. As a second check, we consider the prediction intervals generated by the best performing realized variance forecasts, which is helpful for quantifying forecast uncertainty. We find that the forecasts provide very good covering across prices in all delivery hours.

Finally, we investigate risk premia in electricity forward (futures) markets, where we find that the inclusion of our RCV forecasts vastly improves standard risk premium forecasts based on the model of \textcite{BessembinderLemmon2002}. 

The rest of the paper is structured as follows. In Section~\ref{sec:model}, we introduce the modeling framework and the RCV estimator. In Section~\ref{sec:forecasting_setup}, we introduce the overall matrix-HAR setting and introduce the concrete model specifications to be tested in Section~\ref{sec:models}. Section~\ref{sec:horserace} contains the empirical model horserace along with additional robustness checks, and Section~\ref{sec:risk_premia} investigates the prediction of spread risk premia in forward markets. Section~\ref{sec:conclusion} concludes.

\section{The model and volatility estimation}\label{sec:model}
We follow \textcite{KlosterBenth2026} and impose that the latent price process $X_t(\cdot)$ takes values in the Hilbert space $L^2(\mathbb{S}^1)$, where $\mathbb{S}^1$ denotes the unit circle. The choice of the unit circle was originally proposed in \textcite{Kloster2026} and provides a natural mapping of the daily delivery periods, which respects the periodic nature of the day. Specifically, any time of the day can be represented uniquely by an angle $h\in [0,2\pi)$ via the parametrization $(\cos (h),\sin(h))$, and hence any delivery period $H_i$ can be represented as an interval in $[0,2\pi)$. For example, if delivery period $1$ is from 00:00 to 00:59, then $H_1=[0,2\pi/24)$. We impose that $X_t(\cdot)$ evolves according to the general dynamics 
\begin{equation}\label{eq:latent_price}
dX_t = \mu_tdt + \mathcal{A}X_tdt + \sigma_tdW_t,
\end{equation}
where $W_t$ is a cylindrical Wiener process on $L^{2}(\mathbb{S}^1)$ and the drift $\mu_t$ and volatility $\sigma_t$ are suitably regular processes. In particular, the volatility process $\sigma_t$ takes values in the space of Hilbert-Schmidt operators on $L^2(\mathbb{S}^1)$ and the differential operator $\mathcal{A}$ is assumed to generate a $C_0$-semigroup $\mathcal{S}(t)=\exp(\mathcal{A}t)$. Under sufficient regularity, the dynamics \eqref{eq:latent_price} correspond to the so-called mild solution
\begin{equation}\label{eq:mild_solution}
X_t = \mathcal{S}(t)X_0 + \int_{0}^{t}\mathcal{S}(t-s)\mu_sds + \int_{0}^t\mathcal{S}(t-s)\sigma_sdW_s,
\end{equation}
where the semigroup $\mathcal{S}(t)$ plays the role of the exponential function in the case of a univariate Ornstein-Uhlenbeck type process and is what generates mean-reversion in the price. We refer to \textcite{DaPratoZabczyk2014} for details on mild solution to stochastic equations of the form \eqref{eq:latent_price}.

\subsection{The realized covariation}
In this Section, we expound on the realized covariation estimator of \textcite{KlosterBenth2026}, and most of the results can be found in their Section~2. Given a series of price observations $(P_{t_n})_{n=1}^{N}$ of the form $P_{t}=\left(P_t^{(1)},\ldots,P_t^{(d)}\right)^\top$ with fixed observation frequency $t_{n}-t_{n-1}=\delta$, it holds due to \eqref{eq:price_local_average} that
\[
P^{(i)}_{t_n} = \langle X_{t_n},g_i\rangle_{L^2(\mathbb{S}^1)}, \quad g_{i}=\frac{1}{(h_{i}-h_{i-1})}\mathbf{1}_{[h_{i-1},h_i)}.
\]
We suppose that the drift $\mu_t$ is predictable and slowly varying such that the price decomposes as
\[
P_{t_n} = m_n + Y_n,
\]
where $Y_n\in\mathbb{R}^d$ is a strictly stationary sequence with mean zero, and $m_n\in\mathbb{R}^d$ is such that
\[
\frac{1}{N}\sum_{n=1}^{N}\lVert m_n -m_{n-1}\rVert^2\xrightarrow[N\to\infty]{a.s.}0.
\]
Then we can estimate the slowly varying trend $m_n$ via a kernel regression of the form
\begin{equation}\label{eq:mean_regression}
(\widehat{a}_n,\widehat{c}_n) = \arg\min_{a,c}\sum_{k=1}^{n-1}K\left( \frac{n-k}{\tau}\right)\lVert \widetilde{X}_k-a-c(k-n)\rVert^2,
\end{equation}
and set $\widehat{m}_n=\widehat{a}_n$. Having de-trended the prices, we are left with the series $(Y_n)_{n=1}^{N}$ and we define the semigroup sample estimator $\widehat{\mathcal{S}}_{\delta,N}$ as
\begin{equation}\label{eq:semigroup_sample_estimator}
\widehat{\mathcal{S}}_{\delta,N}
:= \left(\sum_{n=1}^N Y_n Y_{n-1}^\top\right)
\left(\sum_{n=1}^N Y_{n-1}Y_{n-1}^\top\right)^{-1}.
\end{equation}
Under stationarity and ergodicity of $Y_n$, $\widehat{\mathcal{S}}_{\delta,N}$ in \eqref{eq:semigroup_sample_estimator} is a feasible estimator of the matrix $\mathcal{S}_\delta\in\mathbb{R}^{d\times d}$, which is such that $\mathcal{S}_\delta Y_n$ is the best linear one-step predictor of $Y_{n+1}$ based on $Y_n$. In particular, $\mathcal{S}_\delta$ is a finite dimensional matrix representation of the abstract semigroup $\mathcal{S}$, and $\widehat{\mathcal{S}}_{\delta,N}$ is our estimate of this matrix. Defining the plug-in residuals $\widehat{\varepsilon}_n$ by
\begin{equation}\label{eq:plugin_residuals}
\widehat\varepsilon_n := Y_n-\widehat{\mathcal S}_{\delta,N}Y_{n-1},
\end{equation}
we can define the residual-based feasible and annualized realized covariation over $w$ days by
\begin{equation}\label{eq:RCV_pl}
RCV_{n}^w :=
\frac{1}{\delta w}
\sum_{\ell=1}^w
\big(\widehat\varepsilon_{(n-1)w+\ell}\big)
\big(\widehat\varepsilon_{(n-1)w+\ell}\big)^\top.
\end{equation}
The conditional expectation of the residuals $\widehat{\varepsilon}$ is such that
\begin{equation}\label{eq:plugin_residuals_expectation}
\mathbb{E}\left[ \widehat{\varepsilon}_n\widehat{\varepsilon}_n^\top \mid Y_{n-1} \right] = \mathbb{E}\left[ \varepsilon_n\varepsilon_n^\top \mid Y_{n-1}\right] + (\mathcal{S}_\delta - \widehat{\mathcal{S}}_{\delta,(n-1)})Y_{n-1}Y_{n-1}^{\top}(\mathcal{S}_\delta - \widehat{\mathcal{S}}_{\delta,(n-1)})^\top,
\end{equation}
where $\varepsilon_n=Y_n-\mathcal{S}_{\delta}Y_{n-1}$ are the true residuals, computed with the exact, but unknown, semigroup adjustment. The first term in \eqref{eq:plugin_residuals_expectation} is thus the pure innovation variance, whereas the second term arises from the semigroup estimation error. Hence the RCV \eqref{eq:RCV_pl} consists of an innovation term, plus bias from the estimation of $\mathcal{S}_\delta$. If this estimation bias is zero, then
\[
\frac{1}{wN}\sum_{n=1}^{N}RCV_n^w\xrightarrow[N\to\infty]{a.s.} A\int_{0}^{\delta}\mathcal{S}(\delta-s)\sigma_s\sigma_s^\ast\mathcal{S}(\delta-s)^\ast ds A^\ast,
\]
where $A$ denotes the matrix such that $(Ax)_i=\langle x,g_i\rangle_{L^2(\mathbb{S}^1)}$ and $A^\ast$ its adjoint. More precisely, the long-span limit of the RCV estimator \eqref{eq:RCV_pl} is a finite dimensional projection of the semigroup-weighted integrated variance operator, which we refer to as the \emph{effective variance}. Denoting by $\Sigma_t = A\int_{0}^{t}\mathcal{S}(t-s)\sigma_s\sigma^\ast_s\mathcal{S}(t-s)^\ast dsA^\ast$ the cumulative effective variance up to time $t$, the estimator $RCV_n^w$ is then a noisy proxy of the increment $\Sigma_{t_n}-\Sigma_{t_{n-w}}$.

The estimator \eqref{eq:RCV_pl} is straightforward to compute once the data has been properly de-trended. We follow \textcite{KlosterBenth2026} and choose $K$ in \eqref{eq:mean_regression} as an Epanechnikov kernel with a bandwidth $\tau=0.246$, corresponding to 90 days. We also include day-of-week dummies in the regression to eliminate deterministic weekday specific seasonality effects.

\section{Forecasting target and the baseline model}\label{sec:forecasting_setup}
We focus on the short term forecasting of the annualized weekly RCV, corresponding to $w=7$ and $\delta=1/365$. The aim is to produce forecasts of the realized covariation estimator $RCV_n^{w}$ defined by \eqref{eq:RCV_pl}. Concretely, we denote by $\widehat{RCV}_{n+\ell}^w$ the time $t_n$ forecast of $RCV_{n+\ell}^w$. Note that $RCV_n^w$ is computed based on the realized price information up to time $t_n$, whereas $RCV_{n+\ell}^w$ is only known at time $t_{n+\ell}$. The forecasting target is then $RCV_{n+w}^w$, i.e., the full week-ahead realized covariation.

As a baseline model, we propose a HAR type model in the spirit of \textcite{MatrixHAR}, where the variances and the correlation structure are modelled separately. For any matrix $A\in\mathbb{R}^{d\times d}$, we refer to the trivial decomposition $A=DRD$ where $D=\mathrm{diag}\left(\sqrt{A_{11}},\ldots,\sqrt{A}_{dd}\right)$ and $R=D^{-1}AD^{-1}$, as the \emph{DRD-decomposition}. In particular, the RCV has DRD-decomposition 
\[
RCV_t^w = D_t^wR_t^wD_t^w.
\]
The matrix time series $D_t^w$ then contains the $d\times d$ marginal $w$-day realized volatility, and the matrix time seres $R_t^w$ contains the time-varying conditional cross sectional covariance structure, parametrized by $d(d-1)/2$ unique entries. We then model $D_t$ and $R_t$ separately, while ensuring positive definiteness of $RCV^w_{t}$ at all times. Predicting $\widehat{D}_{t+w}^w$ and $\widehat{R}_{t+w}^w$ separately, we simply reconstruct $\widehat{RCV}_{t+w}^w$ as 
\begin{equation}\label{eq:baseline_HAR_model}
\widehat{RCV}_{t+w}^w = \widehat{D}_{t+w}^w\widehat{R}_{t+w}^w\widehat{D}_{t+w}^w.
\end{equation}

\subsection{Modelling the $D_t$ matrix}\label{sec:modelling_D}
The diagonal matrix $D_t=\mathrm{diag}(RCV_t^w)^{\tfrac{1}{2}}$ contains the time varying integrated volatility of each hourly price. Specializing to the weekly RCV, we model the diagonal matrix as
\begin{equation}\label{eq:log_D_HAR_model}
\log(\widehat{D}_{t+7}^7) = \beta_0 + \beta_d\log(\mathrm{diag}(D_t^1))+\beta_w\log(\mathrm{diag}(D_t^7)) + \beta_m \log(\mathrm{diag}(D_t^{28}))+\beta_q \log(\mathrm{diag}(D_{t}^{91})),
\end{equation}
where $\log$ is interpreted as the elementwise natural logarithm. This corresponds to a HAR type model for the integrated volatility, where each hour shares the same regression parameters. This is in contrast to \textcite{MatrixHAR}, where each diagonal element is associated to its own regression and the reason for pooling the regression is twofold: first, we do not have an abundance of data points in the dataset and second, it is demonstrated in \textcite{KlosterBenth2026} that the integrated variance of hourly electricity prices are highly collinear, with a simple level factor driving about 60\% of the variation in the weekly integrated variances.

Conventionally, HAR type models are specified with periods of 1, 5, and 22 days, corresponding to daily, weekly, and monthly estimated realized variance in trading days. As electricity markets operate every day, regardless of weekends and holidays, we instead choose RCV over periods of 7 and 28 days, reflecting weekly and monthly realized covariation. Furthermore, in order to capture longer memory and seasonal behaviour inherent in electricity markets, we add the quarterly component $D_{t}^{91}$. 

\subsection{Modelling the $R_t$ matrix}\label{sec:modelling_R}
Let $\mathrm{vech}(A)$ denote the half-vectorization of a symmetric matrix $A$. We then impose a scalar HAR type model of the form
\begin{equation}\label{eq:R_matrix_model}
\mathrm{vech}\left(\widehat{R}_{t+7}^7 - \overline{R}\right) = \gamma_w \mathrm{vech}\left( R_{t}^{7}-\overline{R}\right) + \gamma_m \mathrm{vech}\left( R_t^{28}-\overline{R} \right) + \gamma_{q}\mathrm{vech}\left( R_t^{91}-\overline{R} \right),
\end{equation}
where, for each $n$, $\overline{R}$ is the expanding window mean of the realized correlations, i.e.,
\[
\overline{R} = \frac{1}{n}\sum_{k=1}^{n}{R}_{k}^w.
\]
As such, the triple $(\gamma_w,\gamma_m,\gamma_q)$ propagates lagged correlations of all hour pairs uniformly. In this case, we omit daily correlations since a rank-1 outer product of the form $RCV_n^1=\Delta Y_n\Delta Y_n^\top$ carries no informative correlation. 

We note that the model \eqref{eq:R_matrix_model} is unrestricted and as such does not guarantee that $\widehat{R}_{t+7}^{7}$ is a valid correlation matrix. However, it turns out that throughout all of our data, the obtained estimate $\widehat{R}_{t+7}^7$ is always a valid correlation matrix, and as such we do not worry about this. To this end, we note that there are various numerical methods developed to estimate models that intrinsically live on the manifold of correlation matrices, such as those developed by \textcite{ArchakovHansen2021}. Such methods would be theoretically more elegant, but to keep the numerical burden light and the exposition simple, we prefer the model \eqref{eq:R_matrix_model}.

\section{Model specifications}\label{sec:models}
In the following, we describe the forecasting model specifications which are applied to German spot price data in Section~\ref{sec:horserace}. Most models will share the same baseline HAR inspired model \eqref{eq:baseline_HAR_model}, where $\widehat{D}_t^w$ is modelled as described in Section~\ref{sec:modelling_D}, and $\widehat{R}_t^w$ is modelled as described in Section~\ref{sec:modelling_R}. This baseline model is labelled \emph{HAR-DRD}. Below, we outline the central extensions of the baseline model that we shall consider and their given labels. We shall also consider combinations of these extensions whenever applicable, but to keep the exposition brief, we do not explicitly characterize all of the tested combinations. Instead, we hope that the model labelling is sufficiently transparent to describe how the combinations work. 
\begin{itemize}
    \item \textbf{Rolling HAR-DRD:} The baseline log-variance regression \eqref{eq:log_D_HAR_model} is estimated based on a 365-day rolling window of data. The correlation regression \eqref{eq:R_matrix_model} still uses an extending window. The justification for this is that we expect correlation dynamics to be more stable over time, and that this makes the estimated long-run mean $\overline{R}$ more stable.
    \item \textbf{HAR-DRD-P:} The baseline HAR-DRD model with an added price covariate $\overline{P}_t=\frac{1}{d}\sum_{i=1}^dP_t^{(i)}$ in the $\widehat{D}_t^w$ regression \eqref{eq:log_D_HAR_model} of the form $\log(1+\lvert \overline{P}_t\rvert)$. 
    \item \textbf{HAR-DRD-P2:} The baseline HAR-DRD model with linear and quadratic price covariates $\overline{P}_t$ and $\overline{P}_t^2$ entering as $\log(1+\lvert \overline{P}_t\rvert )$ and $\log(1+\lvert \overline{P}_t\rvert)^2$. 
    \item \textbf{HAR-DRD+ren:} The baseline HAR-DRD model with the average generation shares of wind and solar generation over the past week added as covariates in the regression \eqref{eq:log_D_HAR_model}. Note that the generation share is always in the interval $[0,1]$.
    \item \textbf{HAR-DRD+tvSpill:} The baseline HAR-DRD model plus time-varying spillover effects added in the $\widehat{D}_t^w$ regression \eqref{eq:log_D_HAR_model} of the form
    \[
    \begin{aligned}
    \log \left( \widehat{D}_{t+7}^w \right) &= (\beta_0 + \beta_0v_t)+ \left(\beta_d+\beta_{d,v}v_t\right)\log(\mathrm{diag}(D_t^1)) + (\beta_w+\beta_{w,v}v_t)\log(\mathrm{diag}(D_t^7))\\
    &\quad +  (\beta_m+\beta_{m,v}v_t) \log(\mathrm{diag}(D_t^{28}))+(\beta_q+\beta_{q,v}v_t) \log(\mathrm{diag}(D_{t}^{91})),
    \end{aligned}
    \]
    where $v_t$ is a scalar regime-dependent variable defined from past observations as
    \[
    v_t = \log\left( \frac{1}{d}\mathrm{Tr}\left(RCV_{t}^{1}\right) \right)-\overline{v},
    \]
    and $\overline{v}$ is the estimated long-run mean of $v_t$. 
\end{itemize}
In addition to these models, we add a daily exponential moving average model as a benchmark of the form
\begin{equation}\label{eq:EWMA_model}
\widehat{RCV}_{t+7}^7 = 7\left(\lambda \widehat{RCV}_{t-1}^1 + (1-\lambda)RCV_{t}^1\right),
\end{equation}
where we select $\lambda=0.94$, corresponding to the standard RiskMetrics methodology. We label the model \eqref{eq:EWMA_model} by EWMA and its inclusion serves to illustrate how well the various HAR-DRD based models compare to a simple non-calibrated benchmark. We also include a plain HAR-DRD model without the quarterly component in \eqref{eq:log_D_HAR_model} and \eqref{eq:R_matrix_model}, to assess whether the quarterly component captures important volatility dynamics. 

\subsection{Loss function and error measures}\label{sec:loss_function}
As is customary in the realized volatility literature \citep[see, e.g.,][]{Patton2011,BollerslevPattonQuadvlieg2016,LaurentRomboutsViolante2013}, we choose a quasi-likelihood loss function $\mathcal{L}$ of the form
\begin{equation}\label{eq:QLIKE_loss}
\mathcal{L} = \log\left(\det\left(\widehat{RCV}_{t+7}^7\right)\right) + \mathrm{Tr}\left( \left(\widehat{RCV}_{t+7}^7\right)^{-1}RCV_{t+7}^7\right).
\end{equation}
The loss function \eqref{eq:QLIKE_loss} penalizes errors asymmetrically; the first term on the right hand side penalizes a forecasted matrix that is ``too large'', while the second term penalizes a forecasted matrix that is ``too small''. However, underestimation is penalized more severely than overestimation by the same amount. 

As additional error measures we consider the Frobenius norm, diagonal mean-squared error, off-diagonal mean-squared error, and the Mincer-Zarnowitz $R^2$ statistic, which is the coefficient of determination of the regression
\[
\mathrm{Tr}\left(RCV_{t+7}^7\right) = \alpha + \beta \cdot \mathrm{Tr}\left(\widehat{RCV}_{t+7}^7\right) + \varepsilon_t.
\]
This measures how much of the variation in realized total weekly variance is linearly explained by the forecast, and as such a higher value of $R^2$ is better. We note, however, that the Mincer-Zarnowitz regression only captures variation in the diagonal, i.e., the integrated variances across hours.

\section{Forecasting horserace}\label{sec:horserace}
In this Section, we consider the forecasting performance of various model combinations as described in Section~\ref{sec:models}. The data is spot price data from the German generation zone, publicly available on the ENTSO-e transparency platform. The dataset is the same as considered in \textcite{KlosterBenth2026} and spans from October 1st 2018 to October 1st 2025. On days affected by daylight savings, we impute the data as follows. If the affected day has 23 hours, we let the missing hour have a price given by the average of the adjacent hours. If the affected day has 25 hours, we have two ``overlapping hours''; the prices in these two hours are merged into one price as their average.

The first 365 day's worth of data are used as a burn-in period to estimate the semigroup $\mathcal{S}_\delta$ in \eqref{eq:semigroup_sample_estimator}. The next 200 day's are then used for the initial training of the models and from then on forecasts are produced daily. Both the semigroup and all of the models are then refit on a rolling 28-day basis. As such, the first day on which we produce forecasts is April 19th 2020 and in total we produce 1984 out-of-sample forecasts of the weekly RCV.

\subsection{Forecasting results}
Table~\ref{tab:horserace} contains the results of the forecasting exercise across 13 different model specifications. We report the quasi-likelihood loss function as well as the error measures outlined in Section~\ref{sec:loss_function}. Furthermore, we report the models which are in the model confidence set in the sense of \textcite{HansenLundeNason2011}, which is essentially the set of models that fail to the reject the null hypothesis of ``equal predictive ability''. We compute the p-values based on 21-day block bootstraps, generating 5000 series of 1984 forecasts.  

For robustness, we report both their range statistic, $T_R$, and their max deviation statistic, $T_\mathrm{max}$. These statistics are computed based on different model comparison criteria and have different power against directions of departure from the null. Roughly speaking, the range statistic filters out models when there are clusters of strong and weak models, whereas the max deviation statistic filters out models when there are a few bad models. Based on the model confidence sets, we observe two things.
\begin{enumerate}
    \item Benchmark and rolling-fitted models are strictly worse than the rest. In particular, the inclusion of the quarterly component seems necessary.
    \item Models with renewables included (wind and solar production share) dominate the rest.
\end{enumerate}
Considering also the other error measures, we note that the inclusion of renewables is detrimental to the overall mean-squared loss and that the inclusion of quadratic price controls greatly benefits the Mincer-Zarnowitz $R^2$ statistic. This indicates that the renewables based models inject some bias, but that the price control allows for better tracking the time variation in the conditional variances. The latter aligns well with the finding in \textcite{KlosterBenth2026} that price and volatility exhibits strong state-dependence. 

Overall, the three surviving models achieve a Mincer-Zarnowitz $R^2$ of $40$-$45\%$, indicating that almost half of the variation in the hourly realized variances is well-explained by the forecasts. 

\begin{figure}[t]
    \centering
    \begin{subfigure}[b]{0.32\textwidth}
        \centering
        \footnotesize
        \textbf{Realized RCV}
        \includegraphics[width=\textwidth]{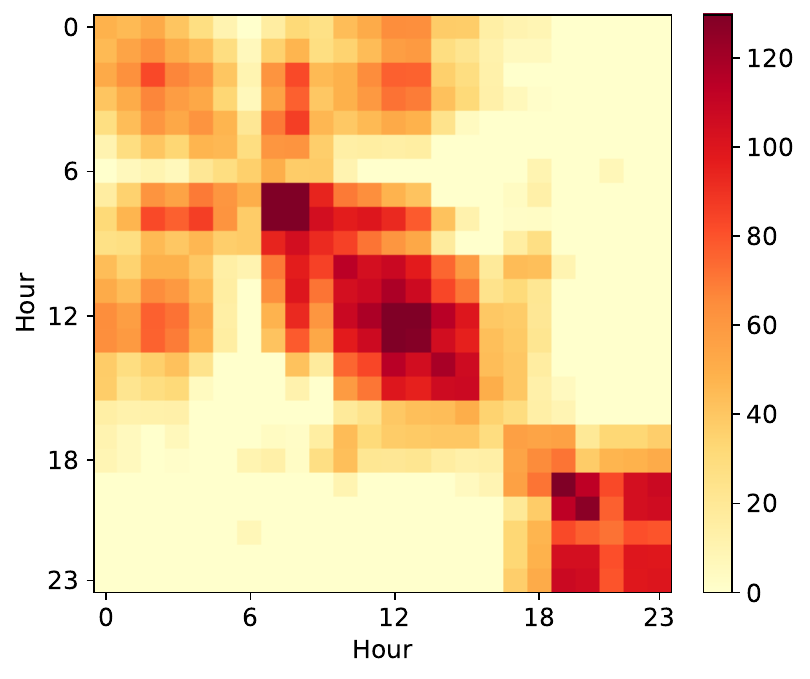}
    \end{subfigure}
    \hfill
    \begin{subfigure}[b]{0.32\textwidth}
        \centering
        \footnotesize
        \textbf{Forecasted RCV, DRD-HAR-P2+ren}
        \includegraphics[width=\textwidth]{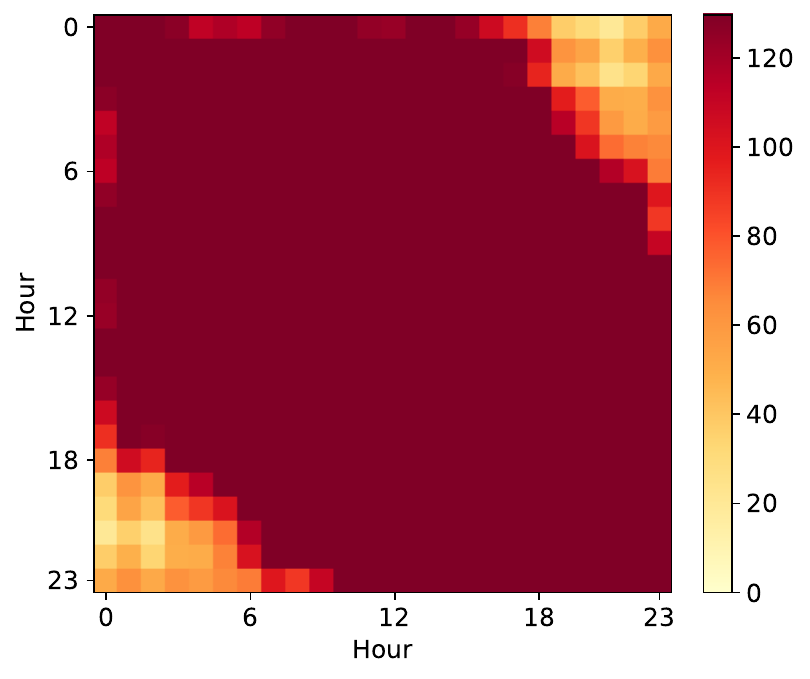}
    \end{subfigure}
    \hfill
    \begin{subfigure}[b]{0.32\textwidth}
        \centering
        \footnotesize
        \textbf{Forecasted RCV, DRD-HAR-P2+ren+tvSpill}
        \includegraphics[width=\textwidth]{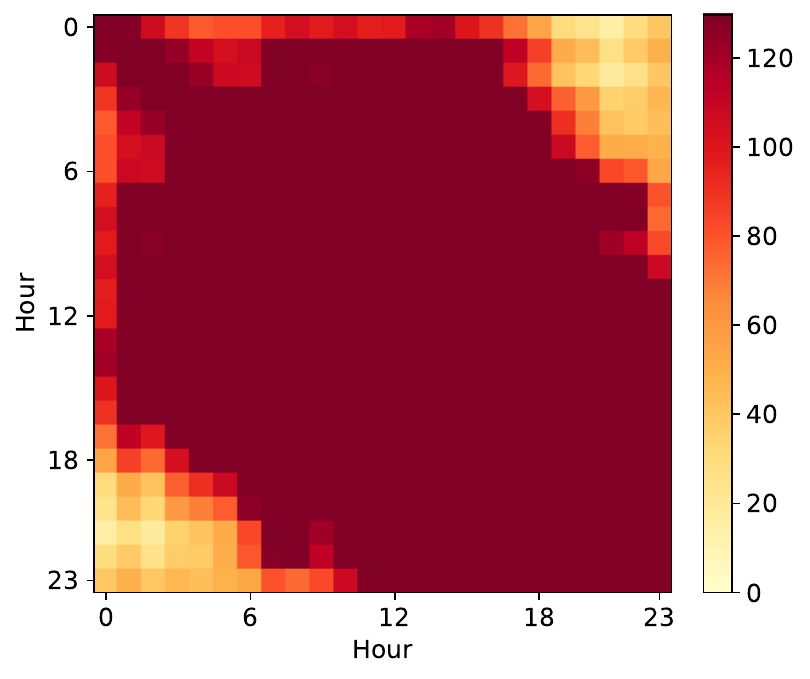}
    \end{subfigure} 
    \vspace{1cm}
    \begin{subfigure}[b]{0.32\textwidth}
        \centering
        \includegraphics[width=\textwidth]{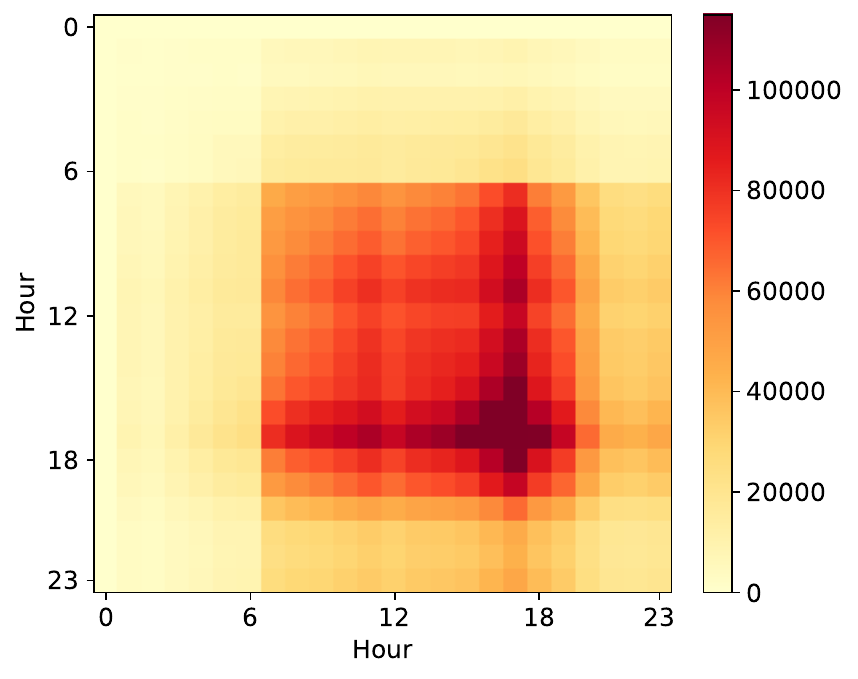}
    \end{subfigure}
    \hfill
    \begin{subfigure}[b]{0.32\textwidth}
        \centering
        \includegraphics[width=\textwidth]{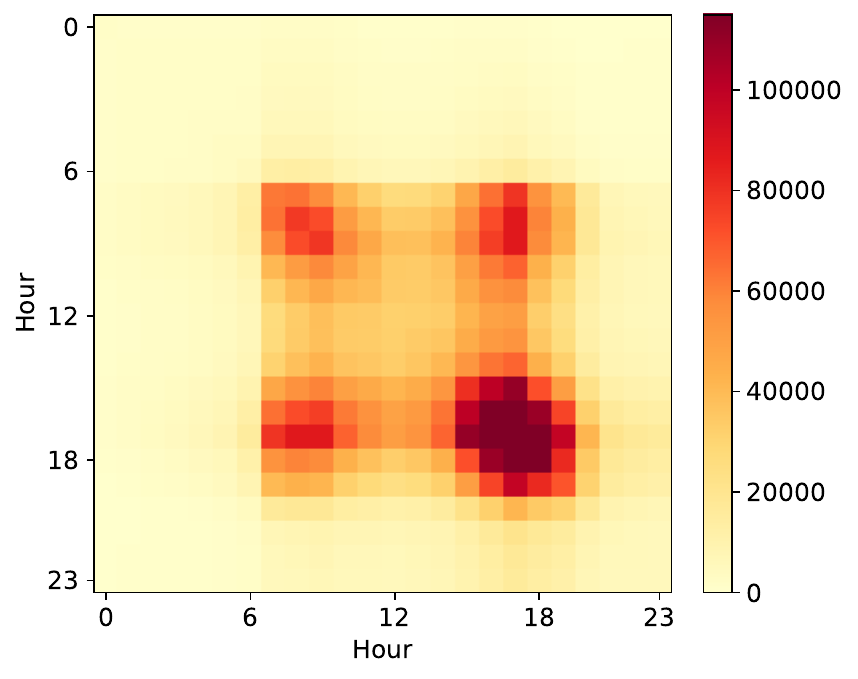}
    \end{subfigure}
    \hfill
    \begin{subfigure}[b]{0.32\textwidth}
        \centering
        \includegraphics[width=\textwidth]{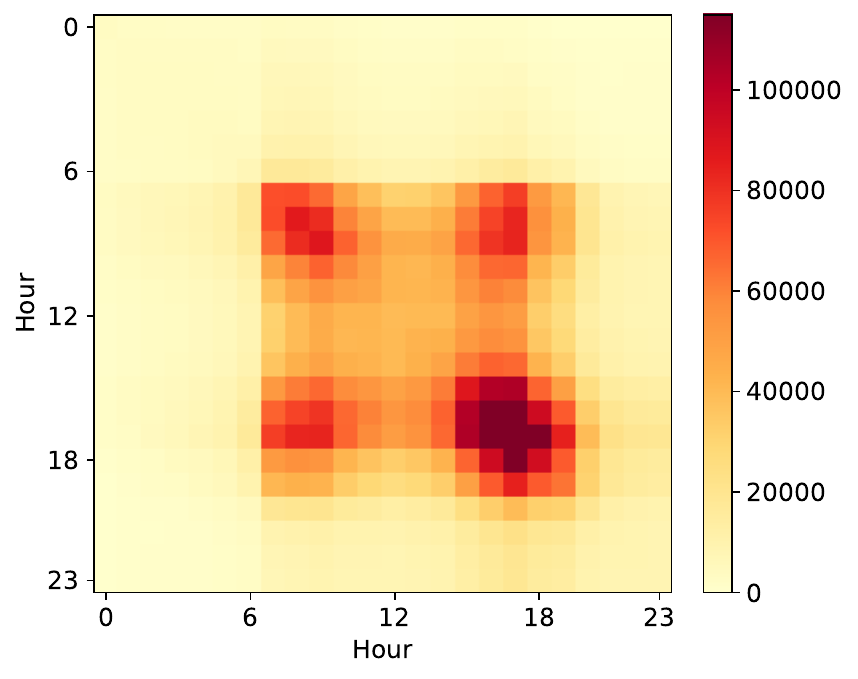}
    \end{subfigure} 
    \caption{Heatmaps depicting the forecasted versus realized RCV matrix on two distinct days for two models. The top panel shows a ``calm'' day (June 19th 2020) while the bottom panel shows a ``turbulent'' day (December 13th 2024). Note that the top and bottom heatmaps are \emph{not} on the same scale.}
    \label{fig:calm_vs_turbulent}
\end{figure}

Figure~\ref{fig:calm_vs_turbulent} depicts heatmaps of the realized and forecasted RCV on two particular days, selected to reflect a ``calm'' and a ``turbulent'' day. It is apparent that the forecasts have a structural floor, likely induced by the right-skewed nature of the volatility on which the models are fitted. The forecasts are therefore essentially unable to meaningfully capture the correlation structure on very calm days. On the turbulent day, where overall volatility is very high, the forecasts are better able to capture the overall clustering structure. These findings are representative of the overall performance and reflects well-known behavior of this type of volatility forecasts.

\begin{figure}
    \centering
    \includegraphics[width=\linewidth]{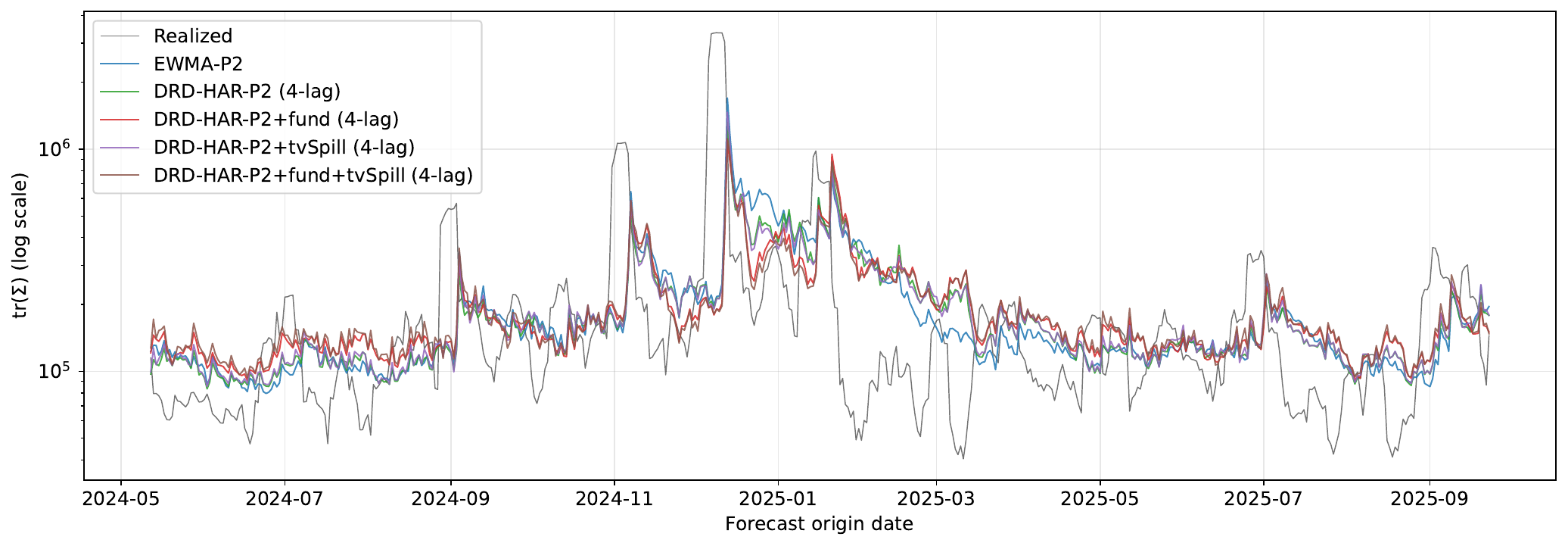}
    \caption{$\mathrm{Tr}\left(\widehat{RCV}\right)$ versus $\mathrm{Tr}\left( RCV\right)$ over time for five of the forecasting models.}\label{fig:trace_timeseries}
\end{figure}

\begin{table}[t]
\centering
\caption{Out-of-sample forecast losses on $RCV_{n+7}^7$ over $n=2191$ days, with $1984$ scored forecast origins (2020-04-19 to 2025-09-23). Losses are mean values across origins. MCS p-values use stationary block bootstrap of length 21 and $5000$ reps. Models marked $\checkmark$ are in the 90\% model confidence set. Bold numbers marks the best in each column.}
\label{tab:horserace}
\scriptsize
\setlength{\tabcolsep}{4pt}
\begin{tabular}{lcrrrrrrcc}
\toprule
\multirow{2}{*}{Model} & \# params & Frob. & {QLIKE} & Diag MSE & Off-diag MSE & {MZ R\textsuperscript{2}} & MCS p & \multicolumn{2}{c}{In MCS} \\
\cmidrule(lr){9-10}
 & & ($\times 10^8$) & & ($\times 10^8$) & ($\times 10^8$) & & ($T_R$-stat) & $T_R$ & $T_\mathrm{max}$ \\
\midrule
\multicolumn{10}{l}{\emph{Benchmarks}} \\
EWMA                              &  0 & 1.399 & 219.78 & 3.550 & 1.306 & 0.3479 & 0.000 &  &  \\
EWMA-P2                           &  4 & 1.338 & 214.33 & 3.334 & 1.251 & 0.4251 & 0.000 &  &  \\
DRD-HAR (3 horizons)    &  6 & 1.309 & 181.08 & 3.434 & 1.217 & 0.3485 & 0.016 &  &  \\
DRD-HAR-P2 (3 horizons) &  8 & 1.270 & 181.00 & 3.341 & 1.180 & 0.4224 & 0.031 &  & $\checkmark$ \\
\midrule
\multicolumn{10}{l}{\emph{4-horizon, expanding window}} \\
DRD-HAR                           &  8 & 1.299 & 180.41 & 3.435 & 1.206 & 0.3734 & 0.031 &  & $\checkmark$ \\
DRD-HAR-P                         &  9 & 1.249 & 180.62 & 3.284 & 1.161 & 0.4060 & 0.031 &  & $\checkmark$ \\
DRD-HAR-P2                        & 10 & 1.263 & 180.47 & 3.343 & 1.172 & \textbf{0.4427} & 0.006 &  & $\checkmark$ \\
DRD-HAR+ren                      & 10 & 1.266 & 179.91 & 3.344 & 1.176 & 0.4056 & 0.168 & $\checkmark$ & $\checkmark$ \\
DRD-HAR-P2+ren                   & 12 & 1.288 & 179.75 & 3.375 & 1.197 & 0.4385 & 0.168 & $\checkmark$ & $\checkmark$ \\
DRD-HAR-P2+tvSpill                & 15 & \textbf{1.235} & 180.42 & \textbf{3.245} & \textbf{1.147} & 0.4339 & 0.016 &  & $\checkmark$ \\
DRD-HAR-P2+ren+tvSpill           & 17 & 1.278 & \textbf{179.65} & 3.339 & 1.189 & 0.4293 & \textbf{1.000} & $\checkmark$ & $\checkmark$ \\
\midrule
\multicolumn{10}{l}{\emph{4-horizon, 365-day rolling window}} \\
DRD-HAR               &  8 & 1.364 & 180.94 & 3.730 & 1.261 & 0.3354 & 0.007 &  &  \\
DRD-HAR-P2          & 10 & 1.341 & 181.00 & 3.790 & 1.235 & 0.3896 & 0.003 &  &  \\
\bottomrule
\end{tabular}
\end{table}

\subsection{A variance-targeting exercise}\label{sec:variance_targeting}
Of the three models in the $T_R$-based model confidence set, the principle of parsimony prescribes that the simple DRD-HAR+ren model with 10 parameters is preferable. In this Section, we consider a risk management exercise to assess how well calibrated the best performing forecasting models are, with particular emphasis on the off-diagonal forecasts. Our overall findings indicate that the full DRD-HAR-P2+ren+tvSpill is competitive, but not significantly better and that excluding the renewables from the model is detrimental.  

To assess the off-diagonal calibration of the models, we consider various ``portfolios'' that reflect meaningful positions/hedges in electricity markets. The motivation is that electricity producers have very different exposure to price and volatility risk, depending on their production capabilities. For example, photovoltaic production will be mostly exposed to mid-day prices and independent of what happens during nighttime, whereas a producer with storage-and-dispatch capabilities such as batteries or power-to-X will have dynamic and varying risk exposures, depending on their strategy. The RCV forecast of the effective variance would be helpful in these contexts to assess the riskiness of certain production commitments or strategies.

\begin{figure}
    \centering
    \includegraphics[width=\linewidth]{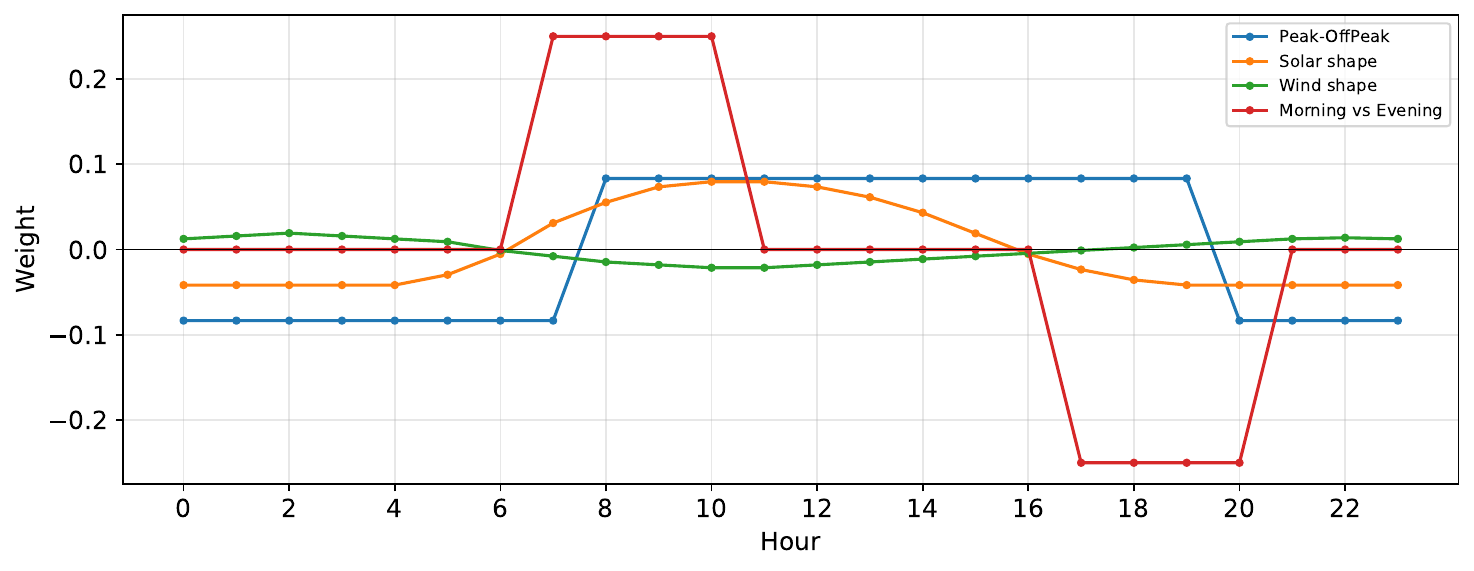}
    \caption{Net-zero volume portfolio profiles.}\label{fig:portfolio_profiles}
\end{figure}

We artificially construct the portfolios to be ``net-zero volume'', such that they reflect pure exposure to price differences between within-day periods (within-day spreads) and a model that correctly forecasts the variance of these spreads is well-calibrated to the within-day cross-correlation structure and not just the marginal variances. We consider the following net-zero-volume positions.
\begin{itemize}
    \item \textbf{Peak vs. off-peak:} A position consisting of equal and opposite positions during peak-hours (hours 8-19) and off-peak-hours (0-7 and 20-23).
    \item \textbf{Solar shape:} A position that mimics the bell shape of solar generation during hours 6-16 and corresponding negative positions in the other hours.
    \item \textbf{Wind shape:} A position that is long in the hours where wind generation tends to be strong (overnight, evening, early morning) and short the mid-day hours.
    \item \textbf{Morning ramp:} A position that is long the four hours 6-9 where prices typically ramp up, and short the hours 2-5, where prices are typically more calm.
\end{itemize}
Figure~\ref{fig:portfolio_profiles} depicts the positions of the four different portfolios. For a vector $\omega\in \mathbb{R}^d$ representing the relative position in the market, we note that the true model implied effective variance of holding the position over the time horizon of one day is given in terms of the effective variance $\Sigma_t$ as
\[
\omega^\top (\Sigma_{t+1}-\Sigma_t)\omega,
\]
and our target is therefore the ratio
\begin{equation}\label{eq:variance_ratio}
VR_t=\frac{\omega^\top RCV_{t+7}^7\omega}{\omega^\top \widehat{RCV}_{t+7}^7\omega}.
\end{equation}
If the forecasting model is well calibrated, we expect ${VR}_t\approx 1$. For each of the best models (and the EWMA baseline) we therefore compute $VR_t$ over time and test if this is overall statistically different from 1, with boostrapped standard errors. We supplement this with a Mincer-Zarnowitz type test of the form
\begin{equation}\label{eq:VR_MZ}
\omega^\top RCV_{t+7}^7\omega = \alpha + \beta\left(\omega^\top\widehat{RCV}_{t+7}^7\omega\right) + \varepsilon_t,
\end{equation}
where $\varepsilon_t$ are zero-mean residuals. The null hypothesis is then that $\alpha=0$ and $\beta=1$. Setting $\widehat{\theta}=(\widehat{\alpha},\widehat{\beta})$, the corresponding Wald test statistic is
$$
\mathcal{W} = (\hat\theta - \theta_0)^\top V^{-1}(\hat\theta - \theta_0), \qquad \theta_0 = (0, 1)^\top,
$$
where $V$ is the estimated HAC-corrected covariance matrix. Under the null hypothesis and standard regularity conditions, we then have $\mathcal{W}\sim \chi^2(2)$.

The results of the experiment are depicted in Table~\ref{tab:variance_ratio_test}, where we see that all three models produce variance ratios \eqref{eq:variance_ratio} that are statistically indistinguishable from 1. This is not the case for the EWMA benchmark and thus indicates that the DRD-HAR based models are learning meaningful volatility dynamics. We note that the full model with price, renewables, and spillover effects included performs best on average, indicating that the increased flexibility might be worthwhile for more delicate risk management tasks. The corresponding Mincer-Zarnowitz regressions based on \eqref{eq:VR_MZ}, however, show that we strongly reject the null of $\widehat{\theta}=(0,1)$ across all models. This is not too surprising as the observed RCV is strongly right-skewed with some very extreme outliers, and our results confirm that the models are not sufficiently reactive during crisis times where volatility spikes. 

\begin{table}[t]
\centering
\caption{Variance-targeting performance and calibration tests on four zero-net-volume spread portfolios, $1984$ scored origins. The 95\% bootstrap CI uses the stationary block bootstrap (block~21, 5000 reps). CIs that contain 1 are marked with $\checkmark$ (mean calibration not rejected at 5\%). MZ Wald test is conducted with Newey-West HAC errors (maxlag 10). Bold marks the mean VR closest to 1 per spread.}
\label{tab:variance_ratio_test}
\footnotesize
\setlength{\tabcolsep}{4pt}
\begin{tabular}{lrlcrr}
\toprule
Model & Mean VR & 95\% boot CI & cal & MZ Wald $p$ & MZ R\textsuperscript{2} \\
\midrule
\multicolumn{6}{l}{\emph{Peak-OffPeak}} \\
EWMA-P2                          & 1.271          & $[1.100,\,1.493]$ & $\times$     & $<\!0.001$ & 0.145 \\
DRD-HAR-P2+ren                  & 1.056          & $[0.870,\,1.325]$ & $\checkmark$ & $<\!0.001$ & 0.144 \\
DRD-HAR-P2+tvSpill               & 1.156          & $[0.968,\,1.415]$ & $\checkmark$ & $\phantom{<}0.001$ & 0.152 \\
DRD-HAR-P2+ren+tvSpill          & \textbf{1.055} & $[0.865,\,1.332]$ & $\checkmark$ & $<\!0.001$ & 0.144 \\
\midrule
\multicolumn{6}{l}{\emph{Solar shape}} \\
EWMA-P2                          & 1.188          & $[1.062,\,1.326]$ & $\times$     & $<\!0.001$ & 0.391 \\
DRD-HAR-P2+ren                  & \textbf{0.994} & $[0.869,\,1.143]$ & $\checkmark$ & $<\!0.001$ & 0.409 \\
DRD-HAR-P2+tvSpill               & 1.087          & $[0.959,\,1.234]$ & $\checkmark$ & $<\!0.001$ & 0.408 \\
DRD-HAR-P2+ren+tvSpill          & 0.981          & $[0.857,\,1.129]$ & $\checkmark$ & $<\!0.001$ & 0.407 \\
\midrule
\multicolumn{6}{l}{\emph{Wind shape}} \\
EWMA-P2                          & 1.201          & $[1.049,\,1.391]$ & $\times$     & $<\!0.001$ & 0.254 \\
DRD-HAR-P2+ren                  & 1.011          & $[0.847,\,1.238]$ & $\checkmark$ & $<\!0.001$ & 0.261 \\
DRD-HAR-P2+tvSpill               & 1.101          & $[0.941,\,1.313]$ & $\checkmark$ & $<\!0.001$ & 0.270 \\
DRD-HAR-P2+ren+tvSpill          & \textbf{0.994} & $[0.832,\,1.218]$ & $\checkmark$ & $<\!0.001$ & 0.261 \\
\midrule
\multicolumn{6}{l}{\emph{Morning vs Evening}} \\
EWMA-P2                          & 1.159          & $[0.998,\,1.355]$ & $\checkmark$ & $<\!0.001$ & 0.268 \\
DRD-HAR-P2+ren                  & \textbf{1.079}          & $[0.897,\,1.287]$ & $\checkmark$ & $<\!0.001$ & 0.255 \\
DRD-HAR-P2+tvSpill               & 1.083          & $[0.897,\,1.302]$ & $\checkmark$ & $<\!0.001$ & 0.268 \\
DRD-HAR-P2+ren+tvSpill          & 1.096          & $[0.915,\,1.304]$ & $\checkmark$ & $<\!0.001$ & 0.256 \\
\bottomrule
\end{tabular}
\end{table}

\subsection{Price forecasting}
The model based on \eqref{eq:latent_price} is conditionally Gaussian. Indeed, conditional on the effective variance $\Sigma_t$, it follows from \eqref{eq:mild_solution} that the time series of observed de-trended prices, $(Y_n)_{n=1}^N$, is conditionally multivariate Gaussian with variance $\Sigma_t$. In particular, the plugin residual $\widehat{\varepsilon}_n$ is approximately conditionally multivariate Gaussian with variance $\Sigma_{t_n}-\Sigma_{t_{n-1}}$, for which the forecasted RCV is an estimate. If the model is properly specified, then the empirical $\alpha$-quantiles of the plugin residuals \eqref{eq:plugin_residuals} should therefore match those predicted by a Gaussian distribution with variance $\widehat{RCV}$. Hence we can, as an additional check of model calibration, check whether the observed residuals $\widehat{\varepsilon}$ lie within the predicted confidence bands based on the forecasted RCV. This is also useful to construct confidence intervals for price forecasts, if we interpret $\int_{0}^{t}\mathcal{S}(t-s)\mu_sds$ as the part which is predicted by a given unspecified model.

\begin{figure}[t]
    \centering
    \includegraphics[width=\linewidth]{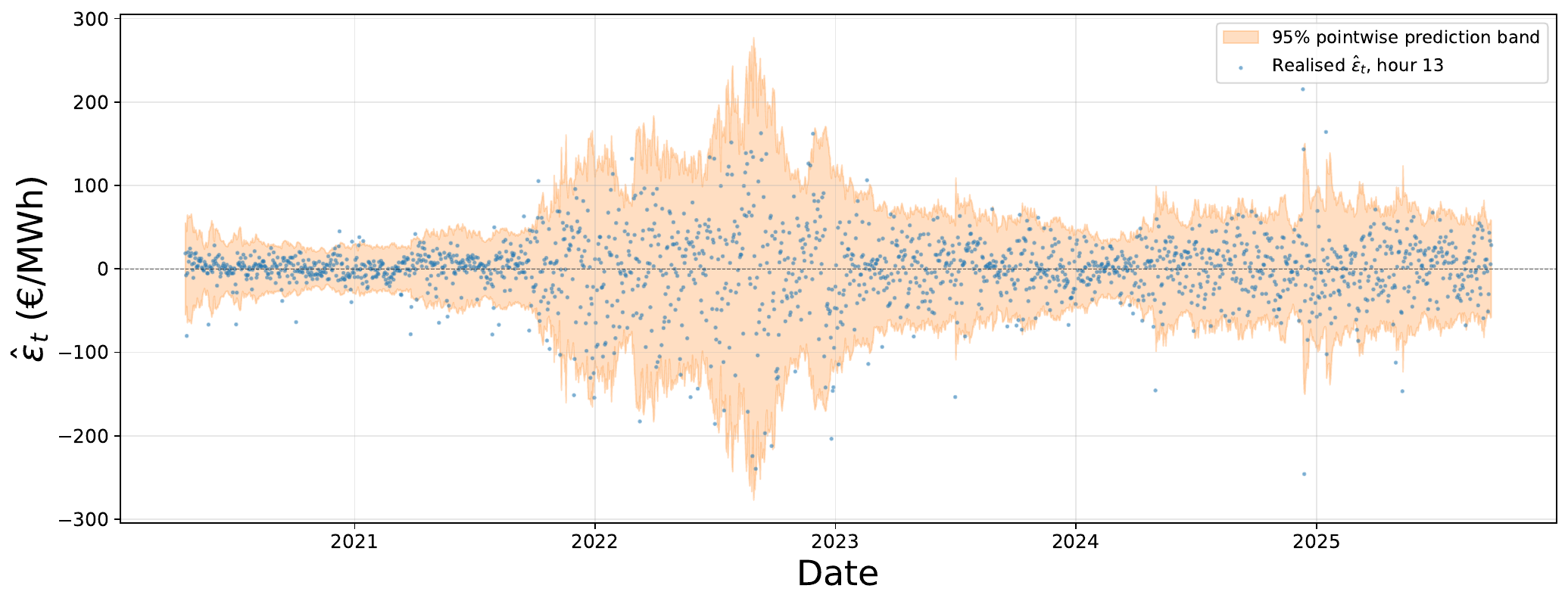}
    \caption{Pointwise 95\% prediction band for delivery period 12-13 over time based on $\widehat{RCV}_{t+7}^7$ predictions from the DRD-HAR-P2+ren+tvSpill model.}
    \label{fig:prediction_band}
\end{figure}

\begin{figure}
    \centering
    \includegraphics[width=\linewidth]{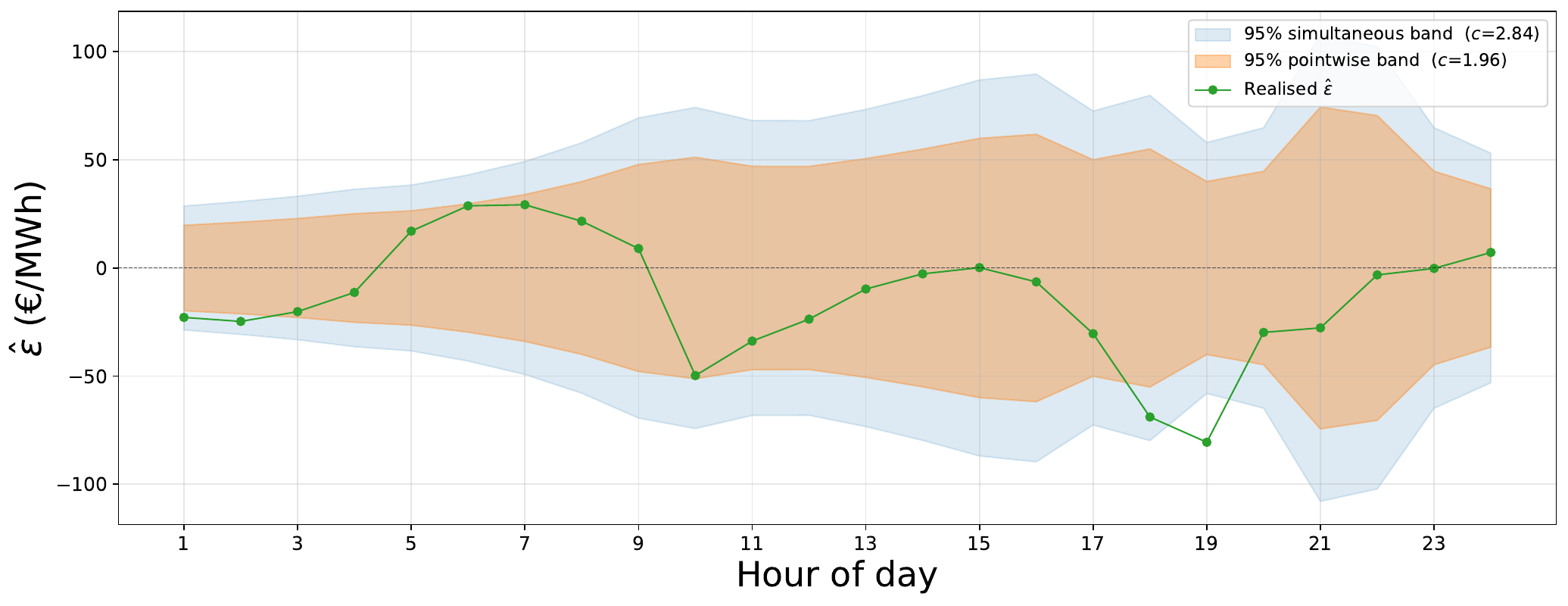}
    \caption{Pointwise versus correlation-adjusted 95\% prediction band for $\widehat{\varepsilon}$ as observed on August 5\textsuperscript{th} 2025 and based on $\widehat{RCV}_{t+7}^7$ predictions from the DRD-HAR-P2+ren+tvSpill model.}
    \label{fig:prediction_band_correlation_adjusted}
\end{figure}

On Figure~\ref{fig:prediction_band}, we depict the univariate time series of residuals $(\widehat{\varepsilon}^{(i)})_{n=1}^{N}$ with $i=13$ (corresponding to the delivery period from 12-13) with a 95\% confidence band superimposed, computed for observation $n$ as $\pm 1.96\widehat{RCV}_{n+7}^7$, i.e., based on the annualized weekly forecast. The forecasting model is the DRD-HAR-P2+ren+tvSpill, which was found to be the best calibrated model on average in Section~\ref{sec:variance_targeting}. Additionally, in Table~\ref{tab:coverage}, we show the achieved coverage from these 95\% prediction intervals across all hours, with an average coverage of 94.7\%. These are computed based on the pointwise multiplier of 1.96, but do not account for the correlation between the entries of $\widehat{\varepsilon}\in\mathbb{R}^d$. To get 95\% confidence bands for the full vector $\widehat{\varepsilon}$, we can bootstrap the correct value $c$ by simulation from a multivariate Gaussian distribution and adjust by $\pm c\cdot\widehat{RCV}_{n+7}^7$. We illustrate the pointwise and multivariate 95\% confidence intervals obtained on a representative date on Figure~\ref{fig:prediction_band_correlation_adjusted} where, on this particular date, the bootstrapped adjustment is $c=2.84$.

\begin{table}[ht]
\centering\small
\caption{Pointwise 95\% prediction-band coverage for the multivariate covariance forecast from the DRD-HAR-P2+ren+tvSpill model.}
\label{tab:coverage}
\begin{tabular}{cc@{\hspace{2.5em}}cc}
\toprule
Hour & Coverage & Hour & Coverage \\
\midrule
1  & 0.965 & 13 & 0.954 \\
2  & 0.963 & 14 & 0.954 \\
3  & 0.960 & 15 & 0.951 \\
4  & 0.951 & 16 & 0.948 \\
5  & 0.954 & 17 & 0.943 \\
6  & 0.954 & 18 & 0.935 \\
7  & 0.950 & 19 & 0.941 \\
8  & 0.947 & 20 & 0.936 \\
9  & 0.945 & 21 & 0.932 \\
10 & 0.951 & 22 & 0.936 \\
11 & 0.953 & 23 & 0.939 \\
12 & 0.951 & 24 & 0.930 \\
\midrule
\multicolumn{4}{c}{\textbf{Overall coverage: 0.947}\quad(nominal: 0.950)} \\
\bottomrule
\end{tabular}
\end{table}

\section{Predicting forward spread risk premia}\label{sec:risk_premia}
It has been demonstrated in the structural equilibrium model of \textcite{BessembinderLemmon2002} that forward market risk premia in electricity markets depend on the variance and skewness of the spot price. Denoting by $F_t(T)$ the time-$t$ price of consuming 1MWh of electricity during day $T\geq t$, they obtain the expression
\[
F_t(T) = \mathbb{E}_t\left[\overline{P}_T\right] + \alpha \mathbb{V}_t\left[ \overline{P}_T \right]+ \beta \mathrm{Skew}_t\left( \overline{P}_T \right),
\]
where $\overline{P}_T$ is the average spot price on day $T$. Defining the forward risk premium $RP_t$ as the difference between forward and expected spot price, we find that
\begin{equation}\label{eq:fwd_risk_premium}
RP_t(T) = F_t(T) - \mathbb{E}_t\left[ \overline{P}_T \right] = \alpha \mathbb{V}_t\left[ \overline{P}_T \right]+ \beta \mathrm{Skew}_t\left( \overline{P}_T \right), 
\end{equation}
which can be estimated via linear regression. In contrast to this approach, it is more common to trade forward contracts with delivery over a longer interval, such as a month, quarter, or year. The risk premium corresponding to \textcite{BessembinderLemmon2002} therefore takes the form
\begin{equation}\label{eq:futures_risk_premium}
    RP_t(T_1,T_2) = \alpha \mathbb{V}_t\left[ \frac{1}{(T_2-T_1)}\sum_{\ell =T_1}^{T_2}\overline{P}_{\ell} \right] + \beta \mathrm{Skew}_t\left[ \frac{1}{(T_2-T_1)}\sum_{\ell =T_1}^{T_2}\overline{P}_{\ell} \right],
\end{equation}
where $[T_1,T_2]$ denotes the delivery period with $t\leq T_1<T_2$. The risk premium \eqref{eq:futures_risk_premium} was investigated by, e.g., \textcite{RedlHaasHuberBohm2009} using data on the EEX exchange. They note that the variance and skewness terms do not admit the same interpretation as in \textcite{BessembinderLemmon2002}, but that they nonetheless may carry useful predictive power. To get good predictions of future risk premia, we thus require good forecasts of the conditional variance of the average price $\overline{P}$, in which case it is natural to use the forecasted realized variance as a proxy. If the underlying of the forward contract is the full daily average price across all delivery periods, we do not require forecasting the full RCV matrix in order to compute the premium \eqref{eq:futures_risk_premium}, as we could simply forecast the realized variance of the univariate average price. However, on the EEX exchange it is possible to trade forward contracts with monthly delivery periods where the underlying reference price is two different averages; \emph{baseload} futures have the full daily average as underlying, whereas \emph{peak-load} futures have only the average of the hours 8-19 as underlying. It is therefore possible to effectively trade a spread between the peak-load and baseload contracts, which leads to a \emph{spread risk premium} of the form
\begin{equation}\label{eq:spread_risk_premium_def}
SRP_t(T_1,T_2) =RP_t^{\mathrm{peak-load}}(T_1,T_2)-RP_t^{\mathrm{baseload}}(T_1,T_2).
\end{equation}
Defining $P^p,P^b$ as shorthands for the scalar-valued total average peak-load and baseload prices over the time interval $[T_1,T_2]$, the spread risk premium corresponding to \eqref{eq:futures_risk_premium} reads
\begin{equation}\label{eq:spread_risk_premium}
    SRP_t(T_1,T_2) = \alpha \mathbb{V}_t\left[ (P^p-P^b)\right] + \beta \mathrm{Skew}_t\left[ (P^p-P^b) \right].
\end{equation}
Since our RCV estimator covers all hours of the day, we can get consistent forecasts of the realized variance of the spread, which proxies the first term in \eqref{eq:spread_risk_premium}. I.e., we identify
\begin{equation}\label{eq:spread_RCV}
    \mathbb{V}_t\left[(P^p-P^b)\right] \approx \mathrm{Tr}\left( \widehat{RCV}_{t+(T_2-t)}^{T_2-T_1} (\omega^p-\omega^b)(\omega^p-\omega^b)^\top \right),
\end{equation}
where the vectors $\omega^b,\omega^p\in\mathbb{R}^d$ are the ``portfolio weights'' of the baseload and peak-load period, such that 
\[
\omega^b=\tfrac{1}{24}(1,1,\ldots ,1)^\top, \quad 
\omega^p = \tfrac{1}{12}(
0,\ldots,0,
\underbrace{1,\ldots,1}_{\text{entries }8,\ldots,19},
0,\ldots,0
)^\top.
\]
The right hand side of \eqref{eq:spread_RCV} is exactly the predicted RCV of a portfolio that is long a peak-load forward contract and short a baseload contract.

\begin{figure}[t]
    \centering
    \begin{subfigure}{0.9\textwidth}
        \centering
        \textbf{Peak-load risk premium}
        \includegraphics[width=0.8\linewidth]{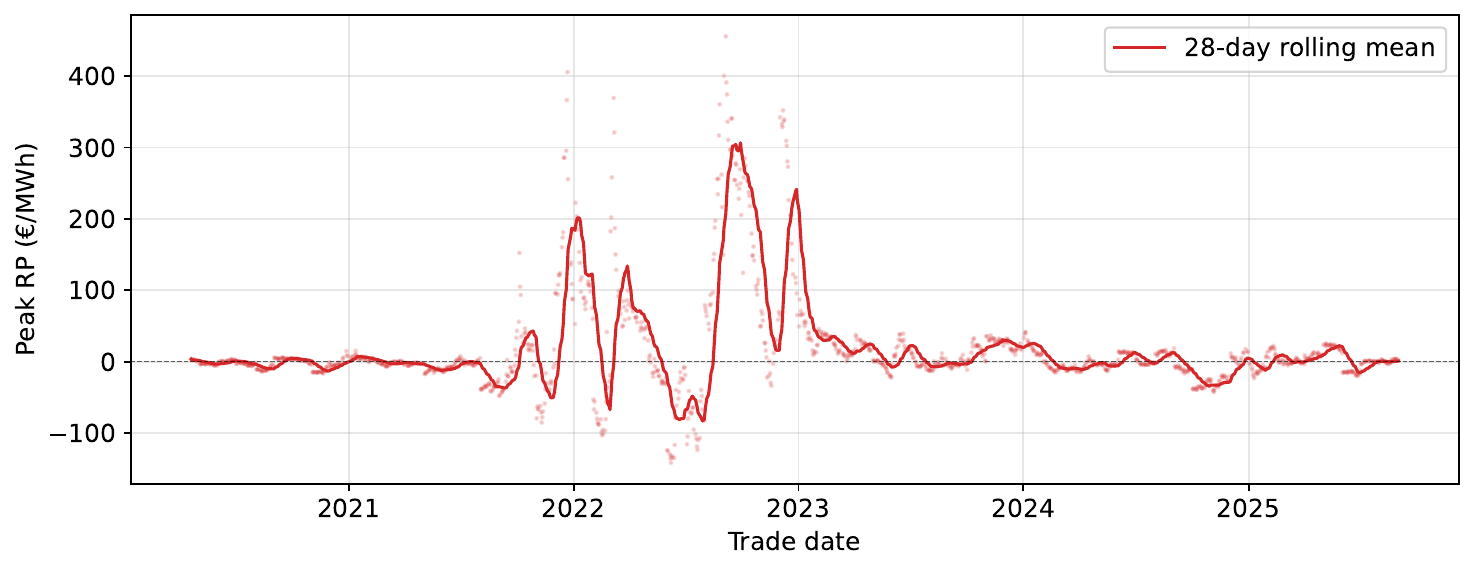}
    \end{subfigure} 
    \hfill

    \begin{subfigure}{0.9\textwidth}
        \centering
        \textbf{Baseload risk premium}
        \includegraphics[width=0.8\linewidth]{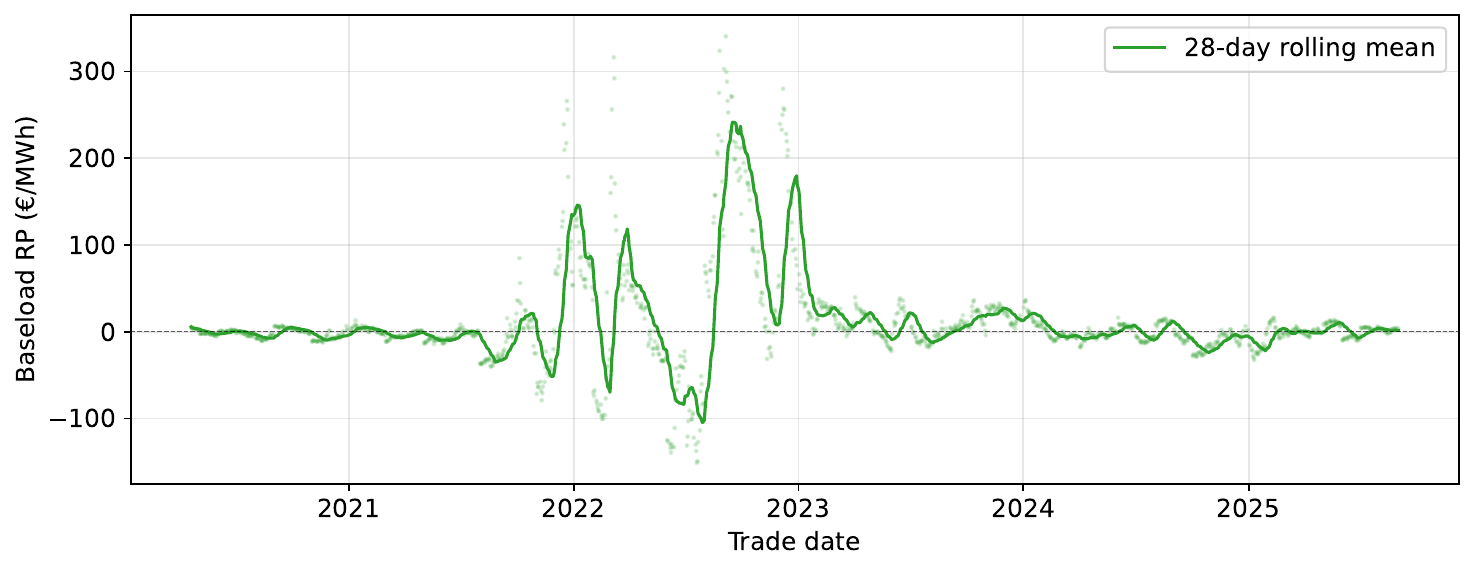}
    \end{subfigure} 
    \hfill

    \begin{subfigure}{0.9\textwidth}
        \centering
        \textbf{Spread risk premium}
        \includegraphics[width=0.8\linewidth]{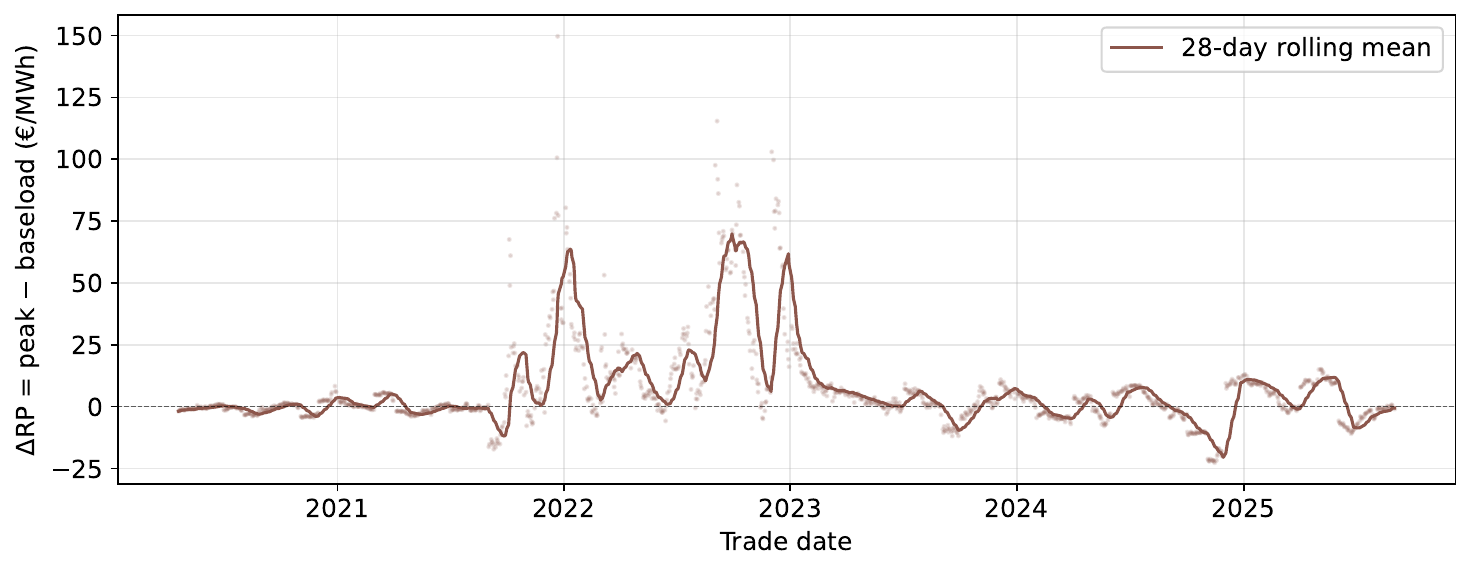}
    \end{subfigure} 
    \hfill
    
    \caption{Estimated risk premia over time.}\label{fig:risk_premia}
\end{figure}

\subsection{An empirical investigation on German data}
We carry out regressions of the form \eqref{eq:spread_risk_premium} in German monthly forward contracts (futures contracts), retrieved from the EEX exchange. The data spans trade dates from March 2nd 2020 to September 30th 2025. The ex-post spread risk premium, $SRP_t$, is estimated at each time $t$ based on the front contract (shortest time to delivery), meaning that $T_1-t$ ranges from 0 to 30 days. On Figure~\ref{fig:risk_premia}, we depict the realized risk premia for baseload and peak-load contracts, as well as the spread risk premium. These are computed ex-post as
\[
\widehat{RP}_t(T_1,T_2) = F_t(T_1,T_2) - \frac{1}{T_2-T_1}\sum_{\ell = T_1}^{T_2}\overline{P}_\ell, 
\]
where $F_t(T_1,T_2)$ is the time-$t$ forward price and $\overline{P}_\ell$ is the realized average price on day $\ell\geq T_1$. The estimated spread risk premium, $\widehat{SRP}_t(T_1,T_2)$, is simply the difference between the estimated baseload and peak-load risk premia, according to \eqref{eq:spread_risk_premium_def}. On Figure~\ref{fig:risk_premia}, we depict the 28-day rolling mean of the estimated risk premia, which reveal a similar time-varying pattern for both peak-load and baseload contracts. Both risk premia spike during the crisis periods of 2022 and 2023, but with the peak-load risk premium being substantially larger in magnitude, thereby leading to a large and positive spread risk premium. Outside of this turbulent period, the spread risk premium tends to fluctuate near zero, with a seemingly seasonal pattern.

In \textcite{BessembinderLemmon2002,RedlHaasHuberBohm2009}, the authors suppose that the conditional variance and skewness entering into \eqref{eq:spread_risk_premium} are approximately equal to their empirical backward looking estimates. We therefore test whether our RCV forecasts carry meaningful information beyond the rolling variance, via the identification \eqref{eq:spread_RCV}. In other words, we set up the regression \eqref{eq:spread_risk_premium} with two different proxies of the conditional future spread variance, given by the backward looking rolling 28-day variance of the spot price spread and the forecasted spread RCV. In both cases, the skewness term is approximated by the backward looking 28-day rolling skewness. We use the weekly RCV forecasts under various different models from Section~\ref{sec:models}. The choice of weekly RCV does not align with the monthly delivery period of the futures, but the annualization in \eqref{eq:RCV_pl} combined with the strong persistence in the RCV means that we expect $RCV^{28}_t$ to be close to $RCV_t^7$ on most days such that, approximately
\[
\mathbb{E}_t\left[\widehat{RCV}_{t+(T_2-t)}^{T_2-T_1}\right] \propto \widehat{RCV}_{t+7}^{7}.
\]
We note that when conducting regressions with the daily risk premia, many observations share the same front month forward contract, such that $F_t(T_1,T_2),F_{t+1}(T_1,T_2),\ldots, F_{T_1}(T_1,T_2)$ are written on to the same underlying average price. This induces heterogeneity in the regression \eqref{eq:futures_risk_premium} and we therefore compute Newey-West HAC-adjusted standard errors with a bandwidth of 60 days in the corresponding regressions. In Table~\ref{tab:spread_rp}, we depict the results of the regression \eqref{eq:spread_risk_premium} for various RCV forecasting models. We note that the estimated $\widehat{\alpha}$ is highly statistically significant in all cases and that $\widehat{\beta}$ is not significant at the 10\% level in any case. The baseline regression with rolling variances yield an $R^2$ statistic of $24.4\%$, while the RCV based models range between $35\%$ and $48\%$, with the best performing model being based on the DRD-HAR-P2 model. In all cases, we find that $\widehat{\alpha}>0$, meaning that the predicted peak-load risk premium is generally higher than the baseload risk-premium. 

\begin{figure}[t]
    \centering
    \includegraphics[width=0.8\linewidth]{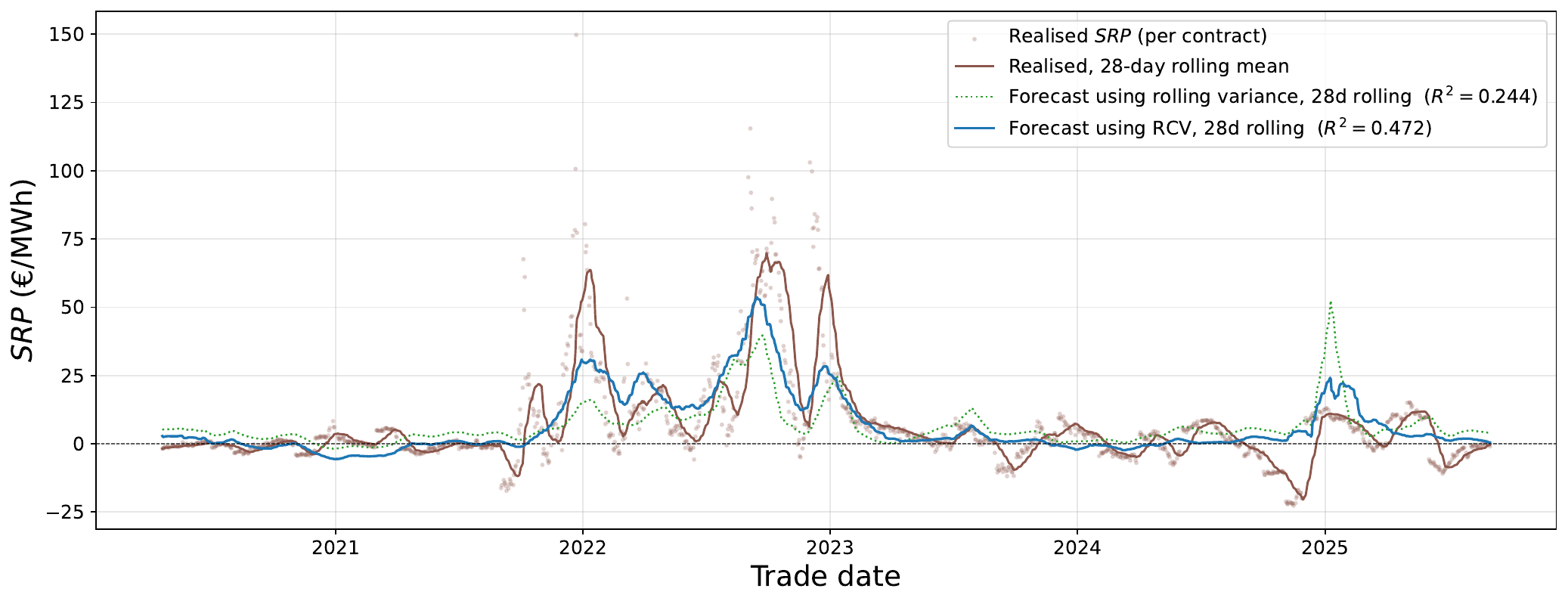}
    \caption{Predicted versus realized spread risk premium using rolling variances and RCV forecasts (DRD-HAR-P2).}\label{fig:predicted_risk_premia}
\end{figure}

On Figure~\ref{fig:predicted_risk_premia}, we depict the spread risk premium forecasts of the baseline regression and the DRD-HAR-P2 based regression. It is apparent that the RCV based model better tracks the shape of the risk premium, in particular in crisis times, but that it is generally difficult to predict when the spread risk premium turns negative. Negative predictions can only arise from the skewness term (since $\widehat{\alpha}>0$), so this indicates that more flexibility is needed in the regression \eqref{eq:spread_risk_premium} or potentially a better proxy for the conditional skewness.

\begin{table}[t]
\centering\small
\caption{Spread risk premium regression \eqref{eq:spread_risk_premium} for various conditional variance proxies. Standard errors are computed by Newey-West HAC correction with a bandwidth of 60 days across $n=1358$ observations.}
\label{tab:spread_rp}
\begin{tabular}{lccccc}
\toprule
Variance proxy & $\hat{\alpha}$ & $\hat{\beta}$ & $R^2$ & $\Delta R^2$ \\
\midrule
\multicolumn{5}{l}{} \\
  Rolling var. (28-day)        & 0.0329$^{***}$ & -1.649 & 0.244 & (baseline)  \\
                                           & (0.0117) & (1.497) &        &           &        \\
  RCV from EWMA-P2                 & 0.0507$^{***}$ & -2.032 & 0.358 & +0.114  \\
                                           & (0.0125) & (1.381) &        &           &        \\
  RCV from DRD-HAR-P2              & 0.0580$^{***}$ & -1.696 & \textbf{0.472} & +0.227  \\
                                           & (0.0093) & (1.233) &        &           &        \\
  RCV from DRD-HAR-P2+ren         & 0.0495$^{***}$ & -0.886 & 0.411 & +0.166 \\
                                           & (0.0094) & (1.178) &        &           &        \\
  RCV from DRD-HAR-P2+tvSpill      & 0.0627$^{***}$ & -1.804 & 0.447 & +0.202  \\
                                           & (0.0112) & (1.257) &        &           &        \\
  RCV from DRD-HAR-P2+ren+tvSpill & 0.0500$^{***}$ & -0.702 & 0.384 & +0.139  \\
                                           & (0.0105) & (1.174) &        &           &        \\
\bottomrule
\multicolumn{6}{l}{\footnotesize Significance: $^{***}\,p<0.01$, $^{**}\,p<0.05$, $^{*}\,p<0.10$.} \\
\end{tabular}
\end{table}

\pagebreak

\section{Conclusion}\label{sec:conclusion}
This paper has demonstrated that the realized covariation of electricity spot prices as introduced in \textcite{KlosterBenth2026} is predictable over short horizons, and that simple matrix-HAR inspired models are well-suited to capture meaningful variation in the conditional second order structure of electricity spot prices. We have found that the inclusion of a quarter-horizon HAR component as well as information on renewable generation share adds important predictive power. These models are well-calibrated to the within-day correlation structure, but suffer from the usual problem of being biased from the extreme right-skew of realized variance. We have also illustrated that using the forecasted realized covariation instead of backward looking variance estimates is useful for predicting risk premia arising in electricity forward markets. The forecast-based predictions are capable of doubling the $R^2$-statistic of the classical risk premium regression of \textcite{BessembinderLemmon2002}.

\section{Disclosure statement}
The authors declare no conflicts of interest.

\section{Funding}
T.~K.~Kloster gratefully acknowledges financial support from the Center of Research in Energy: Economics and Markets and The Danish Council of Independent Research under DFF grant 10.46540/5247-00005B. F.~E.~Benth gratefully acknowledges financial support from the SURE-AI Centre grant 357482, Research Council of Norway. 

\printbibliography

\end{document}